# Twist-angle tunable spin texture in WSe$_2$/graphene van der Waals heterostructures

Haozhe Yang[1,*], Beatriz Martín-García[1,2], Jozef Kimák[3], Eva Schmoranzerová[3], Eoin Dolan[1], Zhendong Chi[1], Marco Gobbi[2,4], Petr Němec[3], Luis E. Hueso[1,2], Fèlix Casanova[1,2,*]

[1] *CIC nanoGUNE BRTA, 20018 Donostia-San Sebastian, Basque Country, Spain.*
[2] *IKERBASQUE, Basque Foundation for Science, 48009 Bilbao, Basque Country, Spain.*
[3] *Faculty of Mathematics and Physics, Charles University, Prague, Czech Republic.*
[4] *Centro de Física de Materiales (CSIC-EHU/UPV) and Materials Physics Center (MPC), 20018 Donostia-San Sebastian, Basque Country, Spain.*

[*]E-mail: haozheyang.nanogune@gmail.com, f.casanova@nanogune.eu

## Abstract

Angle-twisting engineering has emerged as a powerful tool for modulating electronic properties in van der Waals heterostructures. Recent theoretical works have predicted the modulation of spin texture in graphene-based heterostructures by twist angle, although an experimental verification is missing. Here, we demonstrate the tunability of the spin texture and associated spin-charge interconversion with twist angle in WSe$_2$/graphene heterostructures by using spin precession experiments. For specific twist angles, we experimentally detect a spin component radial with the electron's momentum, in addition to the standard orthogonal component. Our results show that the helicity of the spin texture can be reversed by angle twisting, highlighting its critical role on the spin-orbit properties of WSe$_2$/graphene heterostructures and paving the way for the development of novel spin-twistronic devices.

Twisted van der Waals (vdW) heterostructures forming moiré superlattices have garnered significant attention due to their unique, tunable, electronic properties[1–9]. These structures provide an ideal platform to induce and control various physical phenomena, such as magnetic states[4], ferroelectricity[5], and unconventional superconductivity[3,9]. A key ingredient for many of these phenomena is spin-orbit coupling (SOC)[6–8]. Graphene, the archetypal vdW material, exhibits weak SOC, resulting in long spin diffusion length ($\lambda_s$)[10,11] but limited control over spin-dependent phenomena, thus hindering its potential as a platform for spin manipulation. Interestingly, SOC can be enhanced by functionalizing graphene[12–26], which holds great promise for advancing in spintronic applications, such as spin-orbit-based magnetic readout[27] for the magnetoelectric spin-orbit (MESO) logic[28], or potential spin torque applications for magnetic random access memory[29,30].

One representative example of SOC enhancement occurs when graphene is in proximity with a transition metal dichalcogenide (TMD). The strong SOC of the latter introduces the valley-Zeeman ($\lambda_{VZ}$) and Rashba ($\lambda_R$) SOC in graphene, which induce out-of-plane and in-plane spin textures[17,31,32] respectively. The TMD/graphene system was predicted to show spin Hall (SHE)[33–36] and Rashba-Edelstein (REE)[32,35,36] effects. Both of these spin-charge interconversion (SCI) effects were later confirmed experimentally[14,22,24,26]. Recent theoretical works[36–41] predict that the SOC (both $\lambda_{VZ}$ and $\lambda_R$) can be tuned by varying the twist angle between graphene and TMD. At certain twist angles, the breaking of the in-plane mirror symmetry can induce a novel spin texture, leading to unconventional SCI. In a conventional Rashba system, the in-plane mirror symmetry locks the spin of the electron to a direction orthogonal to its momentum, the usual configuration of the REE[42]. If such mirror symmetry is broken, the spin texture gains a radial component, quantified by a Rashba angle $\psi$, which leads to the emergence of an unconventional Rashba-Edelstein effect (UREE)[36,40] with colinear spin-momentum locking. This UREE has been experimentally observed in bulk chiral crystals[43] and metallic TMD/graphene heterostructures[13,20,25,44], without any control of the twist angle. Whereas SCI could thus serve as an important indicator for investigating twist-angle dependent spin texture in moiré heterostructures, no experimental observation has been reported yet.

In this work, we fabricate WSe$_2$/graphene vdW heterostructures with controllable twist angles, which are accurately measured using optical second harmonic generation (SHG) and Raman spectroscopy. SCI was measured with non-local spin precession experiments. Both the REE and the UREE effects are detected in our heterostructures, with an opposite sign when the carriers in graphene transit from electrons to holes and with a maximum near the charge neutrality point of proximitized graphene. Importantly, the UREE shows a sign reversal when twisting the angle, due to the modulation of the in-plane spin texture. This twist-angle-induced UREE is detected up to room temperature. Our findings provide the first experimental demonstration of spin texture modulation by the twist angle, which turns the radial spin component on and off and even changes its sign, leading to a full control of the spin texture helicity. Harnessing this phenomenon, that merges the fields of *spintronics* and *twistronics*, unlocks new opportunities for both fundamental research in moiré heterostructures and practical applications in spin-based devices.

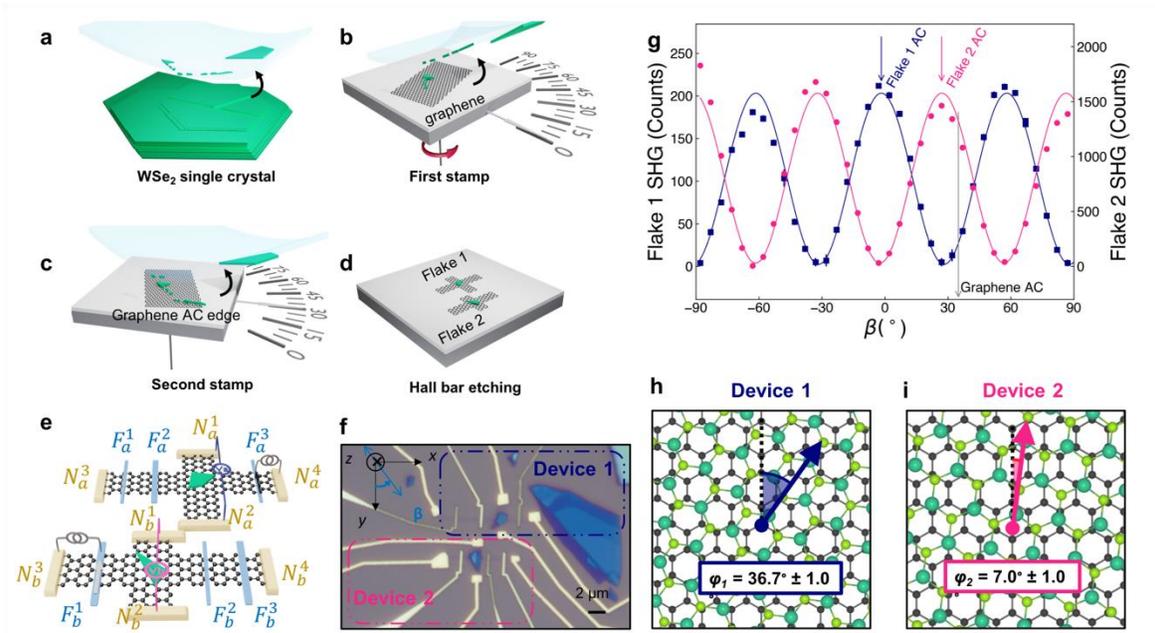

**Fig. 1 Device overview and moiré pattern determination by second harmonic generation. (a)-(e)** Device fabrication process for twisted WSe$_2$/graphene lateral spin valve devices. **(a)** WSe$_2$ flakes exfoliated from the same single crystal onto PDMS. **(b)** The first WSe$_2$ flake is stamped onto a single-layer graphene flake placed on a substrate. **(c)** The substrate is rotated by ~30º, then the second WSe$_2$ flake is stamped onto the same graphene flake. **(d)** The two WSe$_2$ flakes with different crystal orientations on the same graphene are patterned into two individual Hall bars, then **(e)** contacted with non-magnetic contacts (N[1]-N[4]) and ferromagnetic contacts (F[1]-F[3]), corresponding to Devices 1 and 2. Measurement configuration for non-local spin precession experiments is indicated. **(f)** Optical image of Device 1 and Device 2. **(g)** Optical SHG results as a function of the light polarization direction $\beta$ (defined in **(f)**) for the two WSe$_2$ flakes. The armchair (AC) positions are indicated. **(h)** and **(i)** The moiré pattern schematics for Device 1 and Device 2, with the twist angle $\varphi_1 = 36.7° \pm 1.0°$ and $\varphi_2 = 7.0° \pm 1.0°$, respectively.

The device fabrication process is illustrated in Figs. 1a-e. Two WSe$_2$ flakes originating from the same parent single crystal are exfoliated onto a polydimethylsiloxane (PDMS) film, then dry-stamped onto a single-layer graphene flake, exfoliated in a Si/SiO$_2$ substrate, in two steps. In a first step, a WSe$_2$ flake is stamped. In a second step, the substrate is rotated by ~30º and a second WSe$_2$ flake is stamped on the same piece of graphene. These two WSe$_2$/graphene heterostructures with ~30º difference in the twist angle are subsequently patterned into two individual Hall crosses and contacted with several non-magnetic and ferromagnetic (FM) electrodes to form two lateral spin valve devices, as depicted in Fig. 1f. Details of the fabrication process can be found in the *Methods* section, with the optical images during the fabrication process shown in Supplementary Note 1. The two WSe$_2$ flakes keep the same condition after the fabrication process, as confirmed by Raman spectroscopy (Supplementary Note 2).

The crystallographic orientation of the WSe$_2$ flakes is determined by optical SHG characterization, as this non-linear optical process gives the information of the susceptibility coefficients, which reflect the symmetries of the underlying crystal structure[45,46]. The SHG intensity as a function of the polarization direction, $\beta$, of the laser beam for WSe$_2$ Flake 1 and Flake 2 is shown in Fig. 1g, exhibiting a clear sinusoidal behavior with the maximum intensity when the polarization is parallel to the armchair orientation. The SHG intensity of WSe$_2$ (with a six-fold crystal symmetry) behaves as $\cos(3\beta - 3\delta)^2$, where $\delta$ denotes the angle between the armchair axis of the crystal and the y-axis[46]. We observe a 29.3° ± 1.0° difference between Flake 1 and Flake 2, in excellent agreement with the rotation angle during the fabrication process. The crystallographic axis of the single-layer graphene is determined by the edge Raman spectra[47,48], as the armchair edge would contribute to the Raman D-peak (details given in Supplementary Note 2). As a result, the relative crystal orientation between WSe$_2$ and graphene can be precisely determined. We define the twist angle as $\varphi$, where $\varphi = 0°$ signifies that the armchair direction of the graphene is aligned with the armchair direction of the WSe$_2$. Here, the twist angles for Device 1 and Device 2 are $\varphi_1 = 36.7° ± 1.0°$ and $\varphi_2 = 7.0° ± 1.0°$, respectively, as shown in Figs. 1h and 1i.

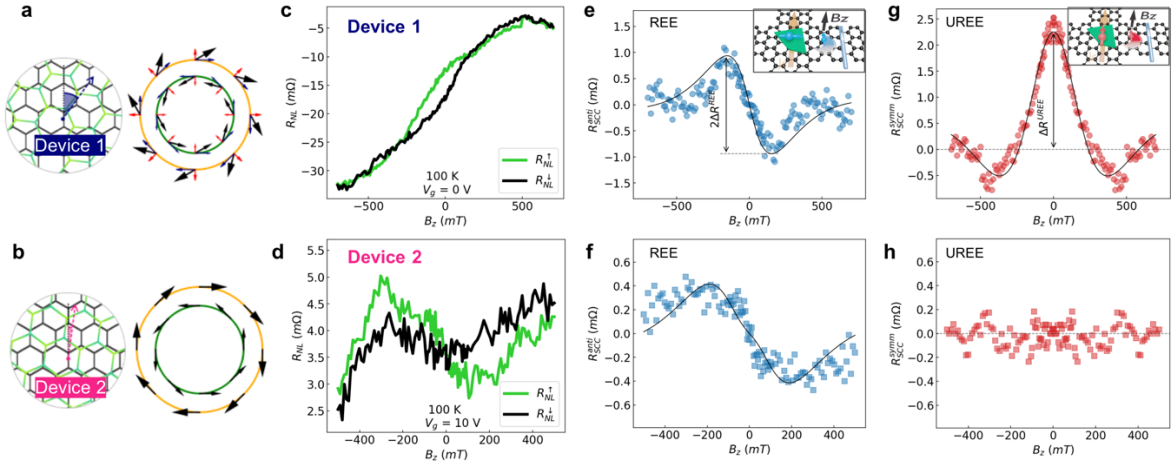

**Fig. 2 In-plane non-local spin precession measurements of the REE and UREE at different twist angles.** **(a,b)** Schematic illustration of the moiré pattern and the in-plane spin texture of **(a)** Device 1 and **(b)** Device 2, with the blue/red arrows standing for the tangential/radial spins. **(c,d)** Non-local spin resistance as a function of $B_z$ using the configuration shown in Fig. 1e, with initial positive ($R_{NL}^\uparrow$, green line) and negative ($R_{NL}^\downarrow$, black line) magnetization direction of the spin injector ($F_a^3$) for **(c)** Device 1, measured at 100 K and $V_g = 0$ V, and **(d)** Device 2, measured at 100 K and $V_g = 10$ V. **(e,f)** Net antisymmetric and **(g,h)** symmetric Hanle precession signals extracted from the two curves in **(c)** and **(d)** by taking $R_{SCI} = (R_{NL}^\uparrow - R_{NL}^\downarrow)/2$ and then extracting the antisymmetric and symmetric components. The black solid lines are a fit of the data to the solution of the Bloch equation. The amplitude of the signals, $\Delta R^{REE}$ and $\Delta R^{UREE}$, are labeled. The antisymmetric and symmetric components originate from REE and UREE, respectively, schematically shown in the insets of **(e)** and **(g)**.

After confirming the twist angle between graphene and WSe$_2$, we studied the spin injection and transport properties of our devices. Initially, we used two FM electrodes as the spin injector ($F_a^1$) and detector ($F_a^2$) and performed Hanle spin precession experiments (see details in Supplementary Note 3). Once we confirmed the spin transport properties of the pristine graphene and the spin polarization of the FM in our device, we performed the SCI measurements. WSe$_2$, as a semiconductor, has a much higher resistance than graphene and, therefore, all charge and spin transport occur through graphene in our heterostructures.

The electrical configuration of the spin-to-charge conversion measurements is illustrated in Fig. 1e. Here, a charge current ($I_c$) is applied from the FM electrode $F_a^3$ to the non-magnetic electrode $N_a^4$. Due to electrical spin injection, a spin current is generated at the interface between $F_a^3$ and graphene, with spin polarization parallel to the FM magnetization along *y*-direction. By applying an out-of-plane magnetic field ($B_z$), the injected *y*-spins precess in the *x-y* plane, causing the diffusing spins to develop an *x*-component at finite magnetic fields. When the spin current reaches the proximitized region, the *x*-spins are converted into a transverse charge current through the REE, resulting in a non-local voltage ($V_{NL}$) between $N_a^1$ and $N_a^2$. The *y*-spins are converted into a charge current through the UREE, since the spin polarization and charge current direction are colinear. To represent our data more accurately, we normalize $V_{NL}$ by $I_c$ to obtain a non-local resistance ($R_{NL}$). Thus, $R_{NL}$ originating from the REE exhibits an antisymmetric Hanle precession behavior, with a maximum at a certain $B_z$ value and a minimum at the same negative $B_z$, whereas the symmetric Hanle precession behavior originating from UREE has a maximum at $B_z = 0$. When reversing the magnetization of the FM electrode, both the symmetric and antisymmetric components of the Hanle precession curve are reversed, because the injector generates the opposite *y*-spin component. Note that charge-to-spin conversion can be measured by swapping the voltage and current terminals, which follows Onsager reciprocity (Supplementary Note 6). For the sake of simplicity, we will refer to both as SCI.

Representative SCI curve sets for Device 1 and Device 2 are shown in Figs. 2c and 2d, respectively. A clear difference is observed between the two sets of curves, which is caused by varying spin textures in the devices, as will be shown below. To obtain the two curves ($R_{NL}^\uparrow$ and $R_{NL}^\downarrow$), we first initialize the $F_a^3$ magnetization with a $B_y$ field and then sweep $B_z$. We repeat this operation for the two $F_a^3$ polarizations and $B_z$ polarities. The net SCI precession signals, $R_{SCI}^{anti}$ from *x*-spins and $R_{SCI}^{symm}$ from *y*-spins, are obtained as follows: Firstly, we take $R_{SCI} = (R_{NL}^\uparrow - R_{NL}^\downarrow)/2$. This step removes any components independent of the initial magnetization, such as local magnetoresistance, ordinary Hall effect, and SCI from the SHE which generates *z*-spins. Secondly, we antisymmetrize and symmetrize the obtained $R_{SCI}$ curve with respect to the magnetic field to separate the *x*-spin (Figs. 2e and 2f) and *y*-spin (Figs. 2g and 2h) contributions. The amplitude $\Delta R^{REE}$ of the antisymmetric curve $R_{SCI}^{anti}$ from REE and the amplitude $\Delta R^{UREE}$ of the symmetric curve $R_{SCI}^{symm}$ from UREE are labeled in Figs. 2e and 2g, respectively.

We observe the antisymmetric precession component in Devices 1 and 2, which indicates the presence of REE (Figs. 2e and 2f, respectively). However, the symmetric component arising from UREE is only observed in Device 1 (Fig. 2g), but not in Device 2 (Fig. 2h). This indicates the different in-plane spin textures for the two devices with different twist angles, as schematically shown in Figs. 2a and 2b. Whereas the broken inversion symmetry in the *z*-direction of the vdW heterostructure yields REE in both devices, the extra broken mirror symmetry in the *x-y* plane due to twisting gives rise to UREE in Device 1. We fitted the net Hanle precession curves with the solution of the Bloch functions in order to extract the SCI efficiency $\alpha_{(U)REE}$ as well as the in-plane spin diffusion length of the proximitized graphene $\lambda_s^\parallel$, with the fitting procedure shown in Supplementary Note 11. For Device 1, $\alpha_{REE} = 0.031\% \pm 0.002\%$, $\alpha_{UREE} = 0.069\% \pm 0.004\%$, and $\lambda_s^\parallel = 380 \pm 120$ nm. For Device 2, $\alpha_{REE} = 0.038\% \pm 0.003\%$ and $\lambda_s^\parallel = 590 \pm 60$ nm. The capability to toggle the UREE on and off is also observed when we switch the charge carriers from holes to electrons, as documented in Supplementary Note 4.

Another mechanism that can give rise to SCI in WSe$_2$/graphene vdW heterostructures is the SHE, where the $z$-spins (out-of-plane) are converted into a transverse charge current. With the same electrical measurement configuration used before, we can now apply the magnetic field along $x$-direction ($B_x$) so that the injected $y$-spins precess in the $y$-$z$ plane, developing a $z$-component at a finite $B_x$. The diffusing spins with a $z$-component are converted into a charge current at the proximitized region giving rise to a non-local voltage antisymmetric with $B_x$ (again normalized into a non-local resistance, $R_{NL}^\uparrow = V_{NL}/I_c$). With this configuration, the conversion of $y$-spins from UREE is also picked up as a symmetric component of $R_{NL}^\uparrow$. When reversing the magnetization of the FM injector, both contributions reverse in the $R_{NL}^\downarrow$ curve. The representative SCI curves set for Device 1, measured at 100 K and $V_g = 20$ V, is shown in Fig. 3a. The net SCI precession signals, $R_{SCI}^{anti}$ from $z$-spins (due to SHE) and $R_{SCI}^{symm}$ from $y$-spins (due to UREE), are obtained as above and plotted in Fig. 3b, with the amplitude of the SHE signal, $\Delta R^{SHE}$, labeled. Note that, in this measurement configuration, the REE-induced SCI of $x$-spins will cause an S-shaped baseline for both $R_{NL}^\uparrow$ and $R_{NL}^\downarrow$, shown with dashed lines in Fig. 3a, due to the pulling of the magnetization of the FM injector along the $x$-axis. In other words, we observe the co-existence of SCI for all three spin components in Device 1. From the fits of the precession curves in Fig. 3b, we obtain the spin Hall angle $\theta_{SHE} = 0.067\% \pm 0.003\%$ and $\alpha_{UREE} = 0.017\% \pm 0.003\%$, with the out-of-plane spin diffusion length of the proximitized region being $\lambda_s^\perp = 550 \pm 70$ nm.

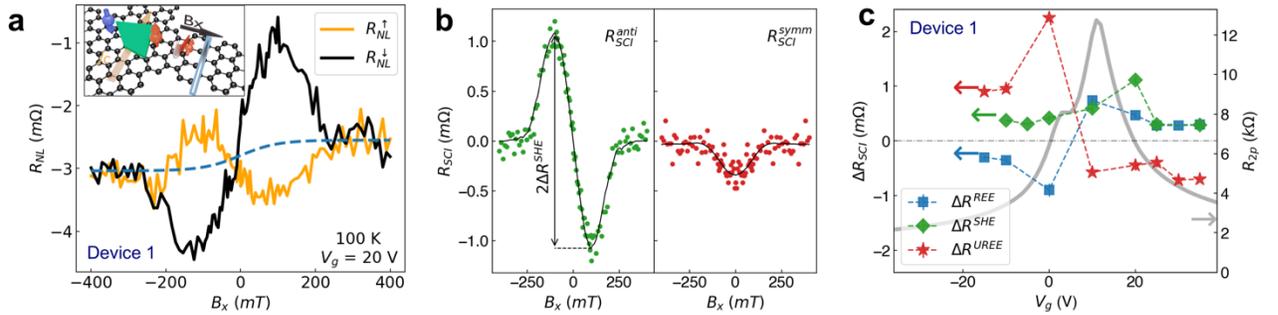

**Fig. 3 Out-of-plane non-local spin precession and gate-dependence of SCI.** (a) Non-local resistance as a function of $B_x$ using the configuration shown in Fig. 1e, with initial positive ($R_{NL}^\uparrow$, yellow line) and negative ($R_{NL}^\downarrow$, black line) magnetization direction of the FM for Device 1 measured at 100 K and $V_g = 20$ V. The inset of (a) shows the schematic of the out-of-plane spin generation originating from the SHE. An S-shaped baseline (dashed lines) is shown, indicating the REE-induced SCI contribution. (b) Net antisymmetric/symmetric Hanle precession signal extracted from the two curves in (a) by taking $R_{SCI} = (R_{NL}^\uparrow - R_{NL}^\downarrow)/2$ and then extracting the antisymmetric and symmetric components, plotted in green/red and originating from the SHE/UREE, respectively. The amplitude of the SHE signal, $\Delta R^{SHE}$, is labeled. **(c)** Gate-dependent SCI amplitude originating from REE (blue squares), SHE (green diamonds), and UREE (red stars). REE and UREE amplitudes are extracted from $B_z$ spin precession, SHE amplitude is extracted from $B_x$ spin precession. The solid gray lines are the two-point resistance of the Hall arm, measured between $N_a^1$ and $N_a^2$ at 100 K.

Next, we studied the gate voltage modulation of the twist-angle-induced SCI. The two-point resistance of the proximitized graphene as a function of the back gate voltage is shown by a solid gray line in Fig. 3c for Device 1. Two peaks from left to right indicate the charge neutral point (CNP) of the proximitized graphene and the pristine graphene, respectively. The gate-dependent charge transport of the pristine graphene is shown in Supplementary Note 3. The SCI amplitudes for the different spin components are plotted in Fig. 3c. Whereas $\Delta R^{REE}$ and $\Delta R^{UREE}$ are extracted from $B_z$ spin precession (Figs. 2e and 2g, respectively), $\Delta R^{SHE}$ is extracted from $B_x$ spin precession (Fig. 3c). One can observe that $\Delta R^{SHE}$ keeps the same sign at all measured gate voltages. $\Delta R^{REE}$, however, changes sign when crossing the CNP. It exhibits the highest value around the CNP, as has also been observed in other TMD/graphene heterostructures[26]. Similarly, $\Delta R^{UREE}$ takes its maximum value around the CNP and changes sign when crossing the CNP. This result is in excellent agreement with the theoretical prediction of the twisted TMD/graphene heterostructure[36], in which changing the carrier type reverses the voltage output sign of the REE and UREE components. An additional experiment that measures the non-local resistance while sweeping $B_y$ is shown in Supplementary Note 5. We replicated the gate-dependence of the SCI with the three spin components (Supplementary Note 9) in an additional sample with twist angle of 39º ± 2º (Device 5, see the fabrication process in Supplementary Note 1), demonstrating the reproducibility of the sign change when crossing the CNP of the proximitized graphene. Interestingly, the SCI conversion stemming from UREE can be detected up to room temperature (Device 5 in Supplementary Note 9).

We conducted experiments with another pair of devices, Device 3 and Device 4, whose twist angles of 41.2º ± 1.7º and 6.5º ± 1.7º are confirmed by SHG (Supplementary Note 10). Again, the UREE can be switched on and off by changing the twist angle between devices 3 and 4 (see Supplementary Note 7). Switching the position of the spin injection FM is detailed in Supplementary Note 8, illustrating a consistent SCI sign when altering the spin current direction for REE and UREE.

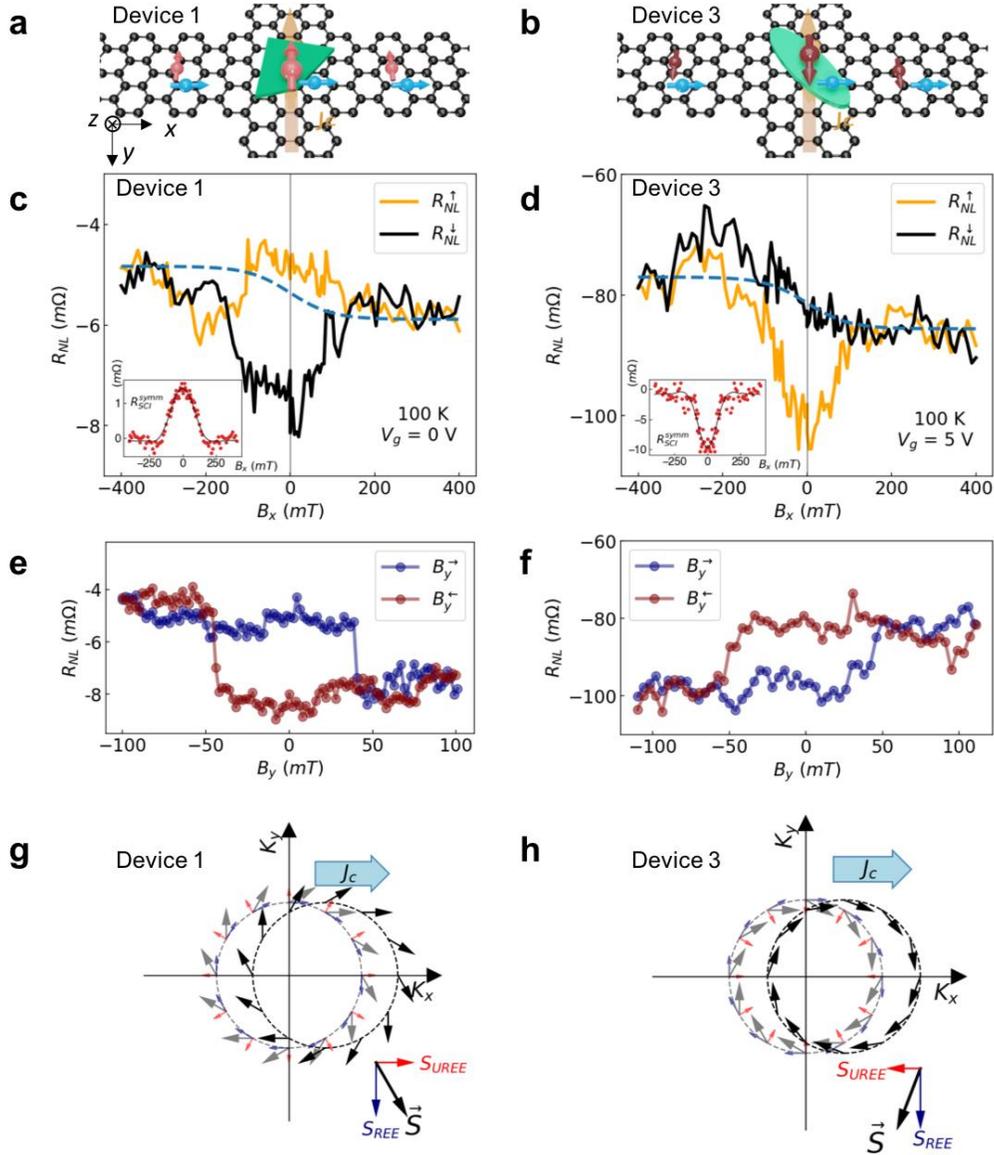

**Fig. 4 Change of helicity in the spin texture with different twist angle.** Sketch of the WSe$_2$/graphene heterostructure with the same REE sign but opposite UREE sign for **(a)** Device 1 and **(b)** Device 3. Non-local resistance as a function of $B_x$ with initially positive ($R_{NL}^{\uparrow}$, yellow line) and negative ($R_{NL}^{\downarrow}$, black line) magnetization direction of FM for **(c)** Device 1 measured at 100 K and $V_g = 0$ V and **(d)** Device 3 measured at 100 K and $V_g = 5$ V. Insets: Net symmetric Hanle precession signal extracted from the two curves in the main panel by taking $R_{SCI} = (R_{NL}^{\uparrow} - R_{NL}^{\downarrow})/2$ and then extracting the symmetric component, exhibiting the opposite UREE sign for the two devices. The S-shaped baseline has the same polarity for the two devices, indicating the same REE sign. Non-local resistance as a function of $B_y$ for **(e)** Device 1 and **(f)** Device 3, showing the UREE signal with opposite polarity. The corresponding spin textures for **(g)** Device 1 and **(h)** Device 3, with only one Fermi contour plotted for clarity.

By comparing devices 1 and 3, we evidence that the UREE signal changes sign by adjusting the twist angle, as illustrated with the sketch of spin accumulation in Figs. 4a and 4b. For a given carrier type (holes in this case), the *x*-spins originating from REE (blue arrows) point in the same direction in both cases. In contrast, the *y*-spins arising from UREE point in opposite direction. We conducted in-plane spin precession experiments, as shown in Fig. 4c and 4d. In agreement with the previous result, the net symmetric Hanle precession components corresponding to the *y*-spins changes sign (see insets). The S-shaped backgrounds, indicated by the blue dashed lines, show the same sign for the SCI originating from *x*-spins (REE) in both Device 1 and Device 3. Indeed, the non-local resistance under a $B_y$ scan, a configuration that picks up exclusively the UREE signal, exhibits a clear sign change between Device 1 (shown in Fig. 4e) and Device 3 (displayed in Fig. 4f). This finding demonstrates that the radial component of the spin texture, originating from the twist-angle-induced in-plane mirror symmetry breaking, reverses at different twist angles. In contrast, the orthogonal component of the spin texture, originating from the out-of-plane inversion symmetry breaking between graphene and WSe$_2$, does not reverse with the twist angle. In other words, we can switch the helicity of the spin texture of WSe$_2$/graphene heterostructures with the twist angle, as sketched in Figs. 4g and 4h.

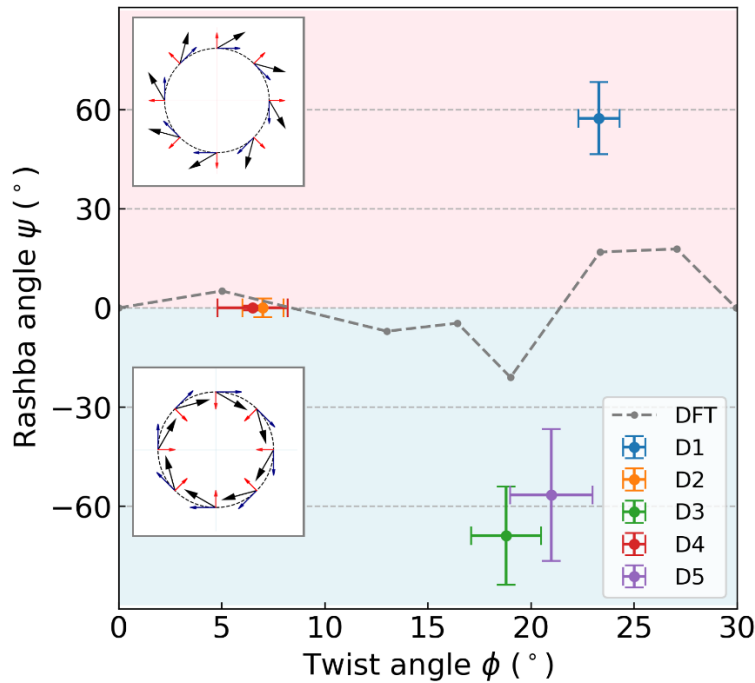

**Fig. 5 Rashba angle tuned by twist angle.** Rashba angle, $\psi$, as a function of the twist angle, $\varphi$, between WSe$_2$ and graphene for Devices 1 to 5. $\psi$ is averaged from the values obtained at different gate voltages and with the corresponding standard deviation shown as the *y*-error bar. The *x*-error bar indicates the fitting error of the SHG for Devices 1-4, and the optical resolution uncertainty for Device 5. The red/blue regions indicate the positive/negative $\psi$. The positive and negative radial component of the spin texture is plotted with only one Fermi contour for clarity. The gray solid dots, with a dashed line, are values taken from the density functional theory (DFT) calculation in ref. [36].

Furthermore, we can quantify the helicity of the spin texture as a function of the twist angle with our SCI measurements. The Rashba angle can be directly calculated by taking $\psi = \arctan(\Delta R^{UREE}/\Delta R^{REE})$. Here, $\Delta R^{UREE}$ and $\Delta R^{REE}$ are extracted from the in-plane $B_x$ SCI measurement to eliminate the amplitude decrease by spin precession. The results are plotted in Fig. 5, showing the Rashba angle as a function of twist angle, averaged from the values obtained at different gate voltages and with the corresponding standard deviation shown as the *y*-error bar. The WSe$_2$ /graphene heterostructure exhibits a 60°-period in-plane rotation symmetry, with an in-plane mirror-symmetry axis at 30° twist angle. Thus, the Rashba angle as a function of twist angle is the same but flips sign between the 0°–30° range and the 30°–60° range of the twist angle, which has been reported in the tight-binding model calculations[40,41]. For the sake of clarity, all devices are represented into the 0°–30° twist-angle range, which required reversing the sign of $\psi$ for Devices 1, 3 and 5, and converting the twist angles of 36.7°, 41.2° and 39° into 23.3°, 18.8° and 21°, respectively. The result shows an excellent agreement with the density functional theory prediction[36], plotted with gray solid dots connected with a dashed line in Fig. 5, with the magnitude and sign varying at different twist angle due to the in-plane symmetry breaking. Whereas the predicted trend in the sign change is very well reproduced, the estimated values of $\psi$ are about three times larger than the prediction. One possibility for such discrepancy can be related to the vertical electric field in the heterostructure, which is predicted to affect $\psi$ (ref. [38]), leading a strong influence to the proximity effect. The tunability of $\psi$ is of fundamental physical interest, but also of interest for potential applications. Further control through a vertical electric field, possibility enabled by a double-gated twisted heterostructure, could potentially be employed to engineer more complex and desirable spin textures for various applications.

To conclude, we experimentally demonstrate the tunability of a spin texture by changing the twist angle between WSe$_2$ and graphene layers. Adjusting the twist angle alters the SOC of the heterostructure, resulting in variations in spin texture that manifest in SCI. A switchable SCI from the UREE originating from the radial spin texture is confirmed by spin precession experiments. This radial spin texture can even change helicity via a slight change of the twist angle. Moreover, the twist-angle-induced UREE remains at room temperature. Our result highlights the ability of tuning spin textures by twist angle, building a link between *spintronics* and *twistronics*. Further exploration of the role of twist angle on SOC is expected to yield valuable insights into the interplay between spintronics and moiré heterostructures, ultimately paving the way for the development of novel spintronic devices with tunable twist angle.

**Methods**

**Sample fabrication:**
Graphene flakes are mechanically exfoliated from natural graphite (NGS Naturgraphit GmbH) onto 300-nm-thick SiO$_2$ on doped Si substrate using Nitto tape (Nitto SPV 224P) and identified using its optical contrast. After selecting graphene flakes, all of them were characterized by micro-Raman spectroscopy to determine the crystal direction of the edge, as shown in Supplementary Note 2.

For Devices 1, 2, and 5, we obtained WSe$_2$ crystals (HQ Graphene) through mechanical exfoliation using Nitto tape and subsequently transferred them onto a piece of polydimethylsiloxane (Gelpak PF GEL film WF 4, 17 mil.). After locating the desired flakes with an optical microscope, we stamped them onto the graphene flake using a viscoelastic stamping tool. This tool employs a three-axis micrometer stage to accurately place the flakes.

The stacking procedure for Devices 3 and 4 can be found in Supplementary Note 1. We mechanically exfoliated WSe$_2$ crystals onto a 300-nm-thick SiO$_2$ layer on a doped Si substrate. We selected a single WSe$_2$ flake with good optical contrast and then used electron-beam lithography to nanopattern it into several elliptic cylinders. Next, we deposited a 20-nm-thick Al layer through vacuum thermal evaporation (base pressure ~1x10$^{-6}$ Torr) to serve as a hard mask. We then reactive-ion etched the WSe$_2$ elliptic cylinders with Ar/SF$_6$ plasma, followed by chemical removal of the Al hard mask using a base developer solution containing tetra-methyl ammonium hydroxide. We picked up one WSe$_2$ elliptic cylinder using a polycarbonate (PC) film, then rotated the Si substrate by a specific twist-angle before picking up the second elliptic cylinder. We stamped both WSe$_2$ elliptic cylinders onto a single graphene flake, forming the heterostructures. Finally, we removed the PC film by dissolving the sample in dichloromethane for 12 h.

After the stamping, the graphene flake is then nanopatterned into Hall bars, also using electron-beam lithography, and covered with a 20-nm-thick Al deposited by high-vacuum thermal evaporation (base pressure ~7×10$^{-7}$ Torr) as a hard mask. The device is then reactive-ion etched with Ar/O$_2$ plasma. The Al hard mask is also chemically removed with a base developer solution with tetra-methyl ammonium hydroxide. The sample is then annealed at 400 °C for 1 hour in an ultrahigh vacuum (10$^{-9}$ Torr) to remove fabrication residues. 45 nm Au was used to contact graphene with 5 nm Pd as seed layer by electron-beam evaporation and lift-off in acetone. The magnetic TiO$_x$/Co electrodes were fabricated by e-beam lithography, deposition of 2.6 Å of Ti (followed by oxidation in the air for 10 minutes), 35 nm of Co, and 20 nm of Au as capping layer by e-beam evaporation, and lift-off. The presence of TiO$_x$ between the Co electrode and the graphene channel leads to interface resistances between 1 and 3 kΩ, measured with a conventional three-point measurement. The widths of the magnetic electrodes are 150 nm or 300 nm in order to have different coercivity.

**Electrical measurements:**
The measurements are performed in a Physical Property Measurement System (PPMS) by Quantum Design, using a DC reversal technique with a Keithley 2182 nanovoltmeter and a 6221 current source. The n-doped Si substrate acts as a back-gate electrode to which we apply the gate voltage with a Keithley 2636B.

**Raman spectra characterization:**
Micro-Raman spectroscopy measurements were carried out at room temperature after the electrical measurement, by placing the samples under the optical microscope of a Renishaw® inVia Qontor Raman instrument equipped with a 100× objective (N.A. 0.85) and a 2400 l/mm grating. For graphene flakes characterization, we used a 532 nm laser as excitation source, while 633 nm laser was used for the WSe$_2$ flakes measurements. The laser power was kept < 1 mW in both cases to avoid the damage of the samples during the measurement.

**Optical Second harmonic generation:**
In order to measure polarization-dependent optical second harmonics (SHG) generation, the sample after all the electrical measurement was excited by a Ti:sapphire oscillator with 150 fs pulse length (Newport Spectra-Physics, Mai Tai, 80MHz repetition rate), with a wavelength fixed to 800 nm. The laser beam was focused by a 20x PLAN APO microscope objective (Olympus LMPLFLN20x, NA 0.4) to the devices, resulting in a spot size below 2 μm. The SHG signal was separated from the fundamental frequency beam by a dichroic mirror (Thorlabs, DMLP550R) and reflected to the detection where, after undergoing additional spectral filtration by two band pass filters (Thorlabs, BG37M), it was focused to the input of a 1 m long bundle of optical fibers. These lead the light to an entrance slit of the spectrograph (Andor, Shamrock 303i, 150 l/mm ruled optical grating with 500 nm blaze) equipped with a CCD camera (Andor, DU420A-OE with 1024×256 pixels) cooled to −65 °C that records the spectra. Polarization of the incoming beam was controlled by a super-achromatic half-waveplate (Thorlabs AHWP10M-580) placed between the dichroic mirror and the objective, which was rotated with a step of 5º. The horizontal projection of the SHG polarization was separated using nanoparticle polarizer (Thorlabs, LPUV050). From the spectra recorded at variable incoming beam polarization, the values of SHG intensity were obtained by a Gaussian fit of the SHG peak.

## Acknowledgements

We acknowledge funding by the "Valleytronics" Intel Science Technology Center, by the Spanish MCIN/AEI/10.13039/501100011033 and by ERDF A way of making Europe (Project No. PID2021-122511OB-I00 and "Maria de Maeztu" Units of Excellence Programme No. CEX2020-001038-M), by the European Union H2020 under the Marie Sklodowska–Curie Actions (Project No. 955671-SPEAR) and by Diputación de Gipuzkoa (Project No. 2021-CIEN-000037-01). Z.C. acknowledges postdoctoral fellowship support from "Juan de la Cierva" Programme by the Spanish MCIN/AEI (grant No. FJC2021-047257-I). B. M.-G. thanks support from "Ramón y Cajal" Programme by the Spanish MCIN/AEI (grant No. RYC2021-034836-I). CzechNanoLab project LM2023051 funded by MEYS CR is gratefully acknowledged for the financial support of the measurements at LNSM Research Infrastructure.

## Author contribution

H.Y. and F. C. conceived the study. H.Y. fabricated the samples, with the help of E.D. and Z.C. H. Y. performed the electrical measurements and the data analysis. B.M.-G and H.Y. performed the Raman spectroscopy measurements. E.S., J. K. and P. N. performed the optical second harmonic generation experiments. All authors contributed to the discussion of the results and their interpretation. H.Y. and F. C. wrote the manuscript with input from all authors.

## Competing interests

The authors declare no competing interests.

## Data and code availability

Source data in the paper are available at https://doi.org/10.6084/m9.figshare.24616287. Any further data and codes used in this study are available from corresponding authors upon reasonable request.

# Supplementary Information

**Note 1 Optical images during the fabrication process.**

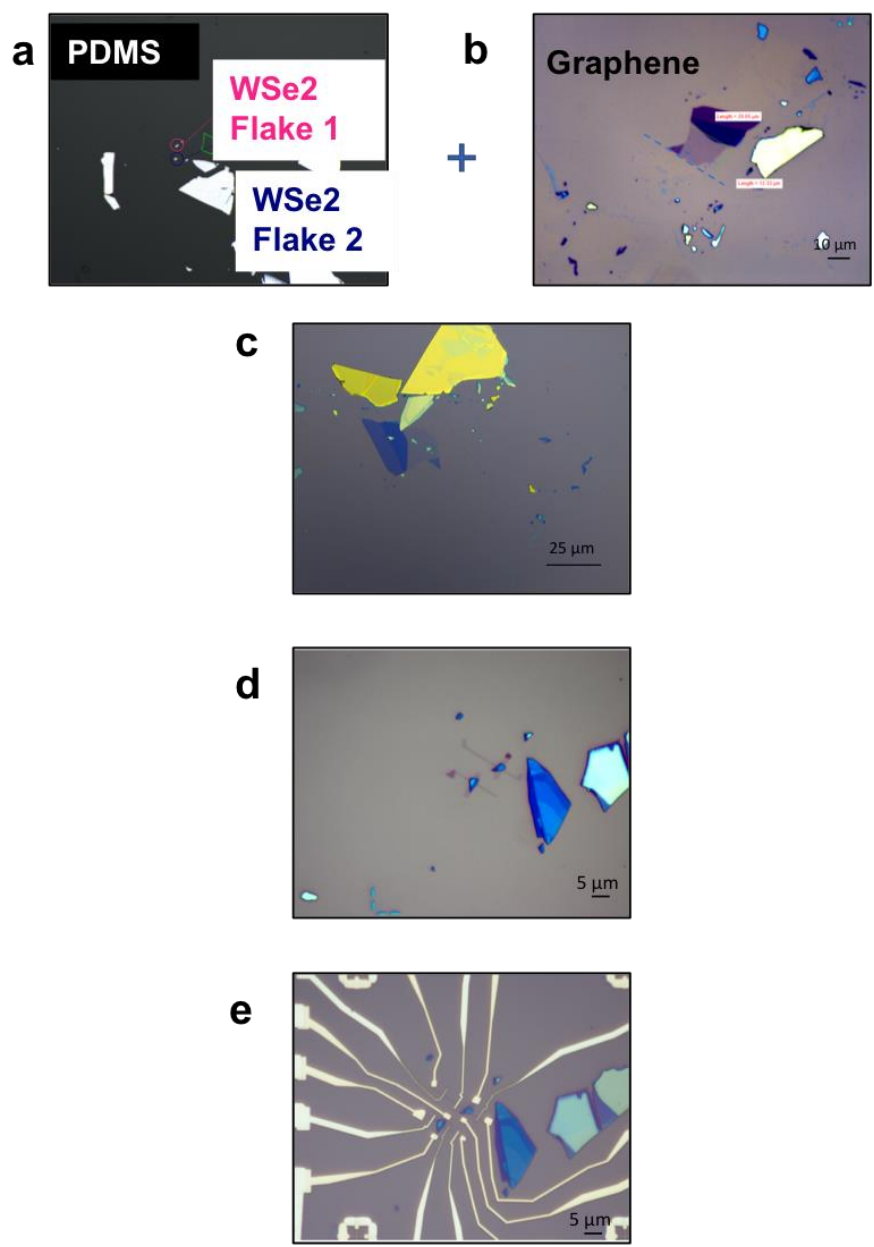

**Fig. S1.1** Optical image during the fabrication process of Device 1 and Device 2. **(a)** Exfoliated $WSe_2$ on PDMS. **(b)** Exfoliated graphene on $SiO_2$/Si substrate, with the blue dashed line indicating the armchair edge identified by Raman spectroscopy. **(c)** The stacked $WSe_2$/graphene heterostructure after two steps of stamping. **(d)** Two Hall bars etched with $WSe_2$ flakes on the cross-junction areas. **(e)** Complete devices with non-magnetic (Ti/Au) and magnetic ($TiO_x$/Co) contacts.

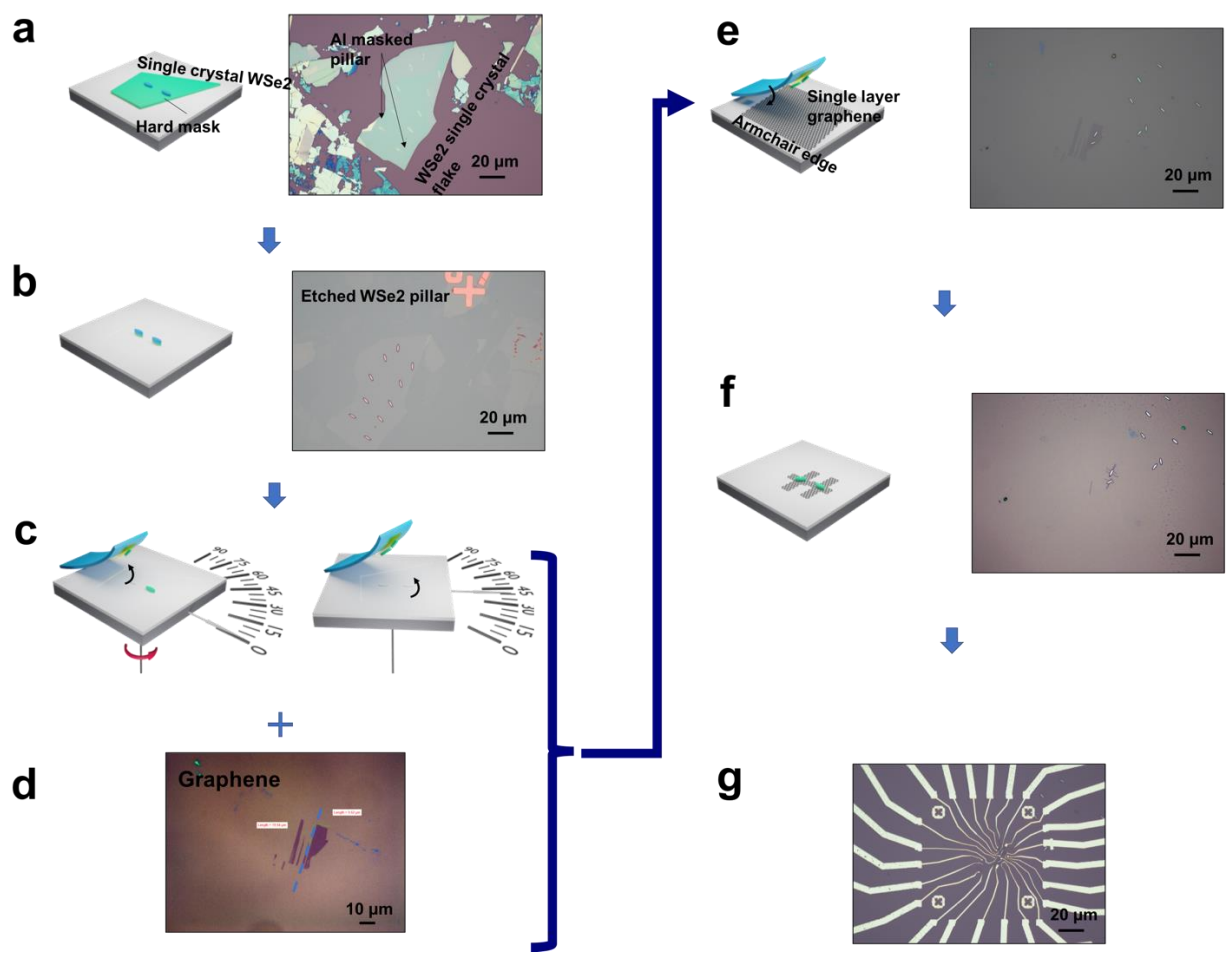

**Fig. S1.2** The schematic fabrication process and the optical images during fabrication step for Device 3 and Device 4. The WSe$_2$ elliptic cylinders are fabricated with hard mask **(a)** and etching **(b)** on a single crystal flake, then picked up by PC membrane on a PDMS film **(c)** and stamped onto a single layer graphene **(d)**, forming two heterostructures **(e)**. The blue dashed line in **(d)** is the armchair edge of graphene confirmed by Raman spectra. Separated Hall crosses are etched with the heterostructure on the cross-section **(f)**. The non-magnetic (Ti/Au) and magnetic (TiO$_x$/Co) contacts are then fabricated, with the devices shown in **(g)**.

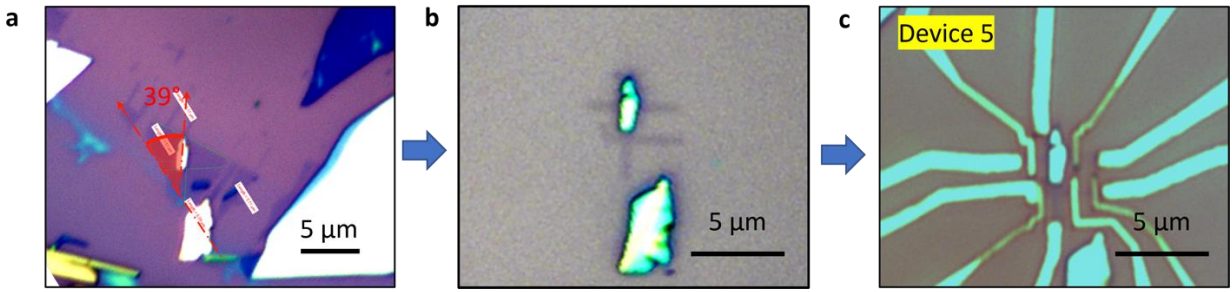

**Fig. S1.3** Optical images during the fabrication process of Device 5. **(a)** One WSe$_2$ flake stamped on the single-layer graphene. The twist angle between graphene and WSe$_2$ is determined with the crystal orientation of the elongated direction as shown. **(b)** The Hall bar after the etching process of graphene. **(c)** The device after fabricating the non-magnetic (Ti/Au) and magnetic (TiO$_x$/Co) contacts.

**Note 2. Raman spectra of the devices.**

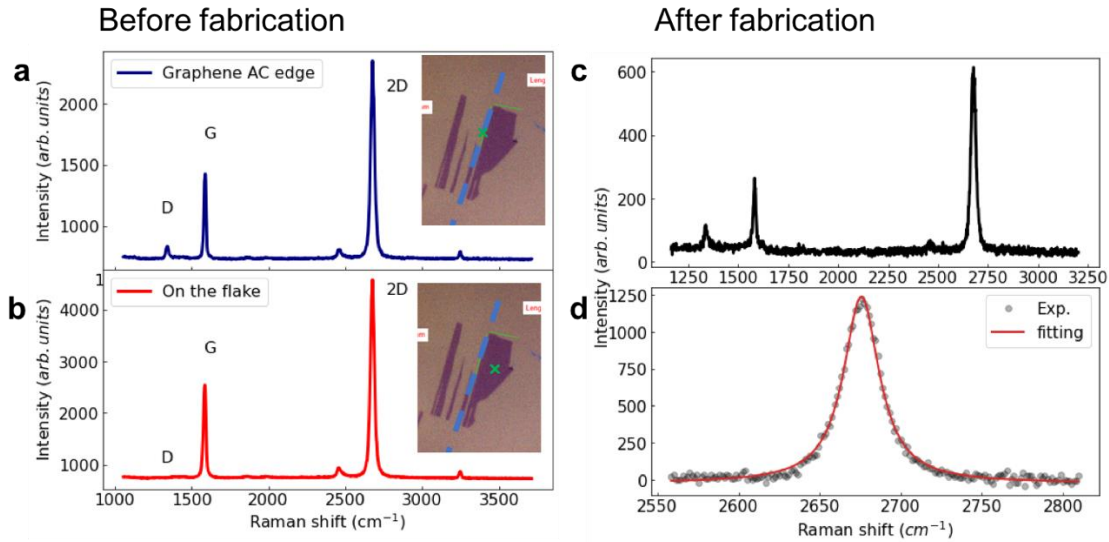

**Fig. S2.1** Representative Raman spectra of the single layer graphene flake used for Device 3 and Device 4, before the fabrication, by locating the laser spot marked with a green cross **(a)** at the edge and **(b)** on the flake. A clear D peak enhancement can be found when the laser is at the edge shown in the inset of (a), as expected for an armchair edge. **(c)** The spectrum after fabrication with the laser spot on the 0.6-μm-wide channel. **(d)** The 2D peak from panel c is fitted with a single Lorentzian function, confirming the layer number of graphene (single layer).

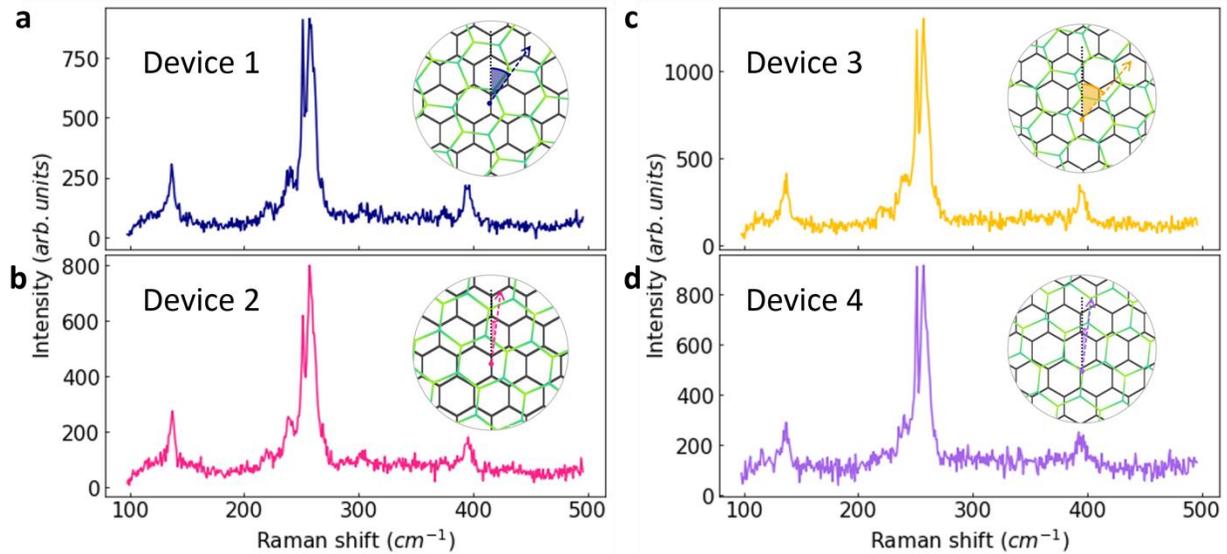

**Fig. S2.2** Raman spectra of the corresponding $WSe_2$ flake after the fabrication process and electrical measurement for **(a)** Device 1, **(b)** Device 2, **(c)** Device 3 and **(d)** Device 4. The $WSe_2$ characteristic peaks are shown clearly and are identical in Raman shift and linewidth for all the flakes, indicating that the properties of the $WSe_2$ are well-preserved for the different twist-angle devices.

**Note 3. Spin injection and transport for pristine graphene.**

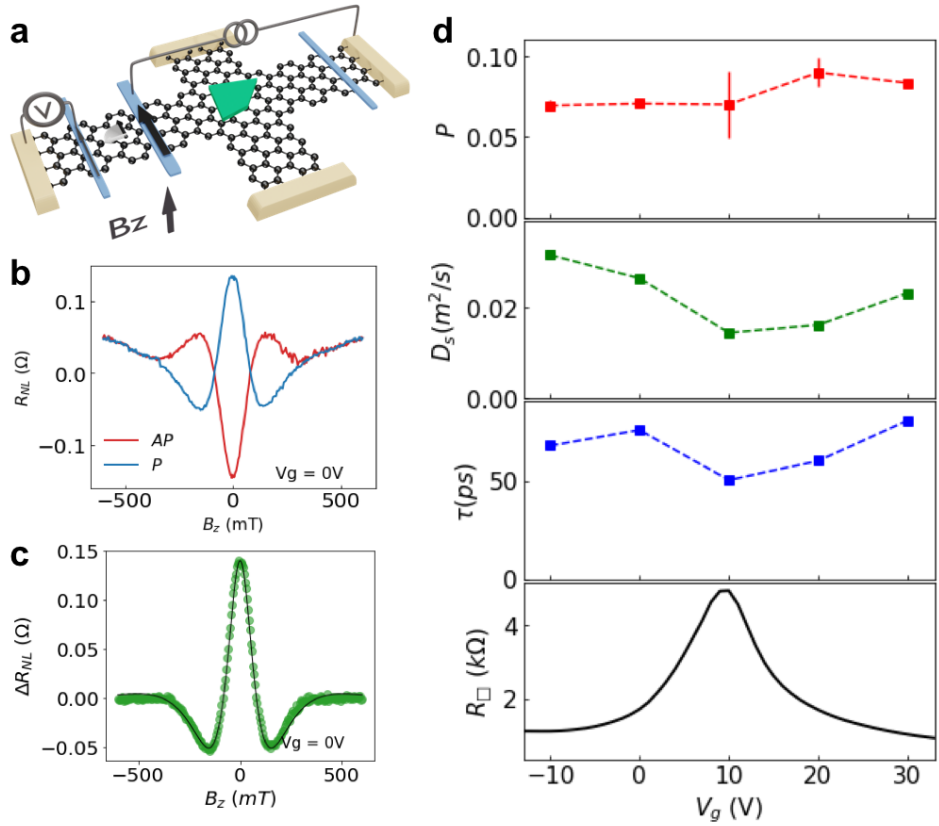

**Fig. S3 (a)** Measurement configuration for spin transport in pristine graphene (symmetric Hanle precession) with standard electrical spin injection and detection using two FM electrodes in Device 1. An out-of-plane magnetic field, $B_z$, is applied to induce precession of the $y$-polarized spins injected from the FM electrode into the graphene. **(b)** Non-local resistance as a function of $B_z$ measured at 100 K and $V_g$ = 0 V using the configuration in **(a)**, with the two FM electrodes set in a parallel ($R_{NL}^P$, blue line) and antiparallel ($R_{NL}^{AP}$, red line) configuration. **(c)** Net symmetric Hanle precession signal extracted from the two curves in **(b)** by taking $\Delta R_{NL} = (R_{NL}^P - R_{NL}^{AP})/2$. The gray solid line is a fit of the data to the solution of the Bloch equation, as explained in Note S11.1. **(d)** The spin polarization ($P$), the spin diffusive constant ($D_s$), and the spin lifetime ($\tau$) of pristine graphene at 100 K, extracted from the fit in panel c, as a function of $V_g$ is plotted with red, green and blue squares. The square resistance, $R_\square$, of the pristine graphene as a function of $V_g$ measured at 100 K using a four-point configuration is shown as a black solid line.

**Note 4. Twist-angle dependent spin-charge interconversion with negative charge carriers.**

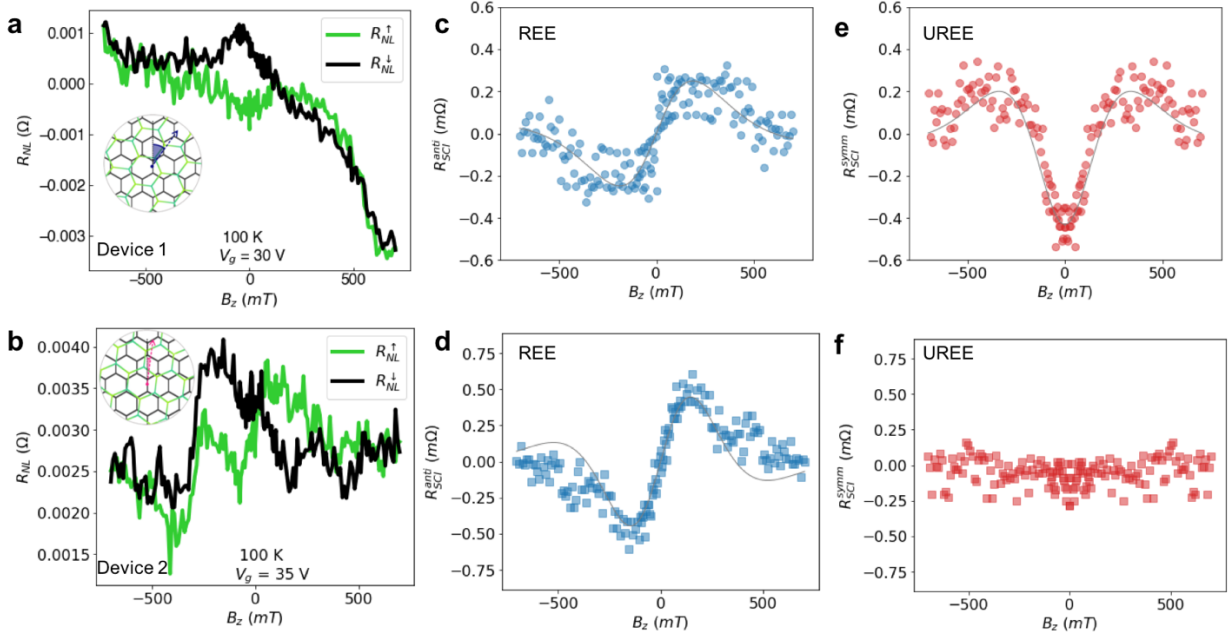

**Fig. S4 (a,b)** Non-local resistance as a function of $B_z$ using the configuration shown in Fig. 1e of the main text, with initial positive ($R_{NL}^\uparrow$, green line) and negative ($R_{NL}^\downarrow$, black line) magnetization direction of the spin injector ($F_a^3$) for **(a)** Device 1, measured at 100 K and $V_g$ = 30 V, and **(b)** Device 2, measured at 100 K and $V_g$ = 35 V. **(c,d)** Net antisymmetric and **(e,f)** symmetric Hanle precession signals extracted from the two curves in **(a)** and **(b)** by taking $R_{SCI} = (R_{NL}^\uparrow - R_{NL}^\downarrow)/2$ and then extracting the antisymmetric and symmetric components, which arise from REE and UREE, respectively. The black solid lines are a fit of the data to the solution of the Bloch equation.

**Note 5. Gate dependence of spin-charge interconversion by sweeping $B_y$ for Device 1.**

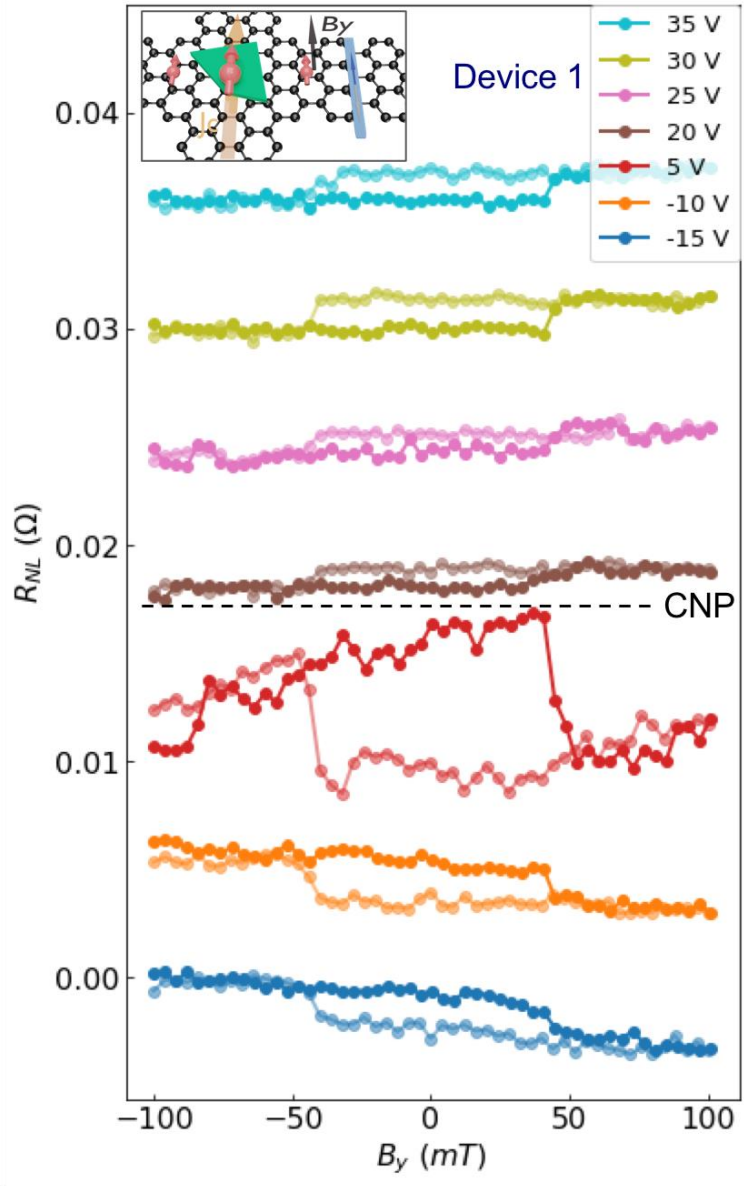

**Fig. S5** Non-local resistance as a function of $B_y$ for Device 1 measured at 100 K and different applied back gate voltage ($V_g$), with the sketch of the measurement shown in the inset. The $y$-spins injected from the FM injector are converted to charge current by the UREE, and the non-local voltage switches sign when the FM switches magnetization. Thus, $R_{NL}$ vs. $B_y$ exhibits the typical easy-axis hysteresis loop. We observe a clear sign change when $V_g$ crosses the CNP, in agreement with the spin precession result in Fig. 3c of the main text.

**Note 6. Onsager reciprocity between spin-to-charge and charge-to-spin conversion.**

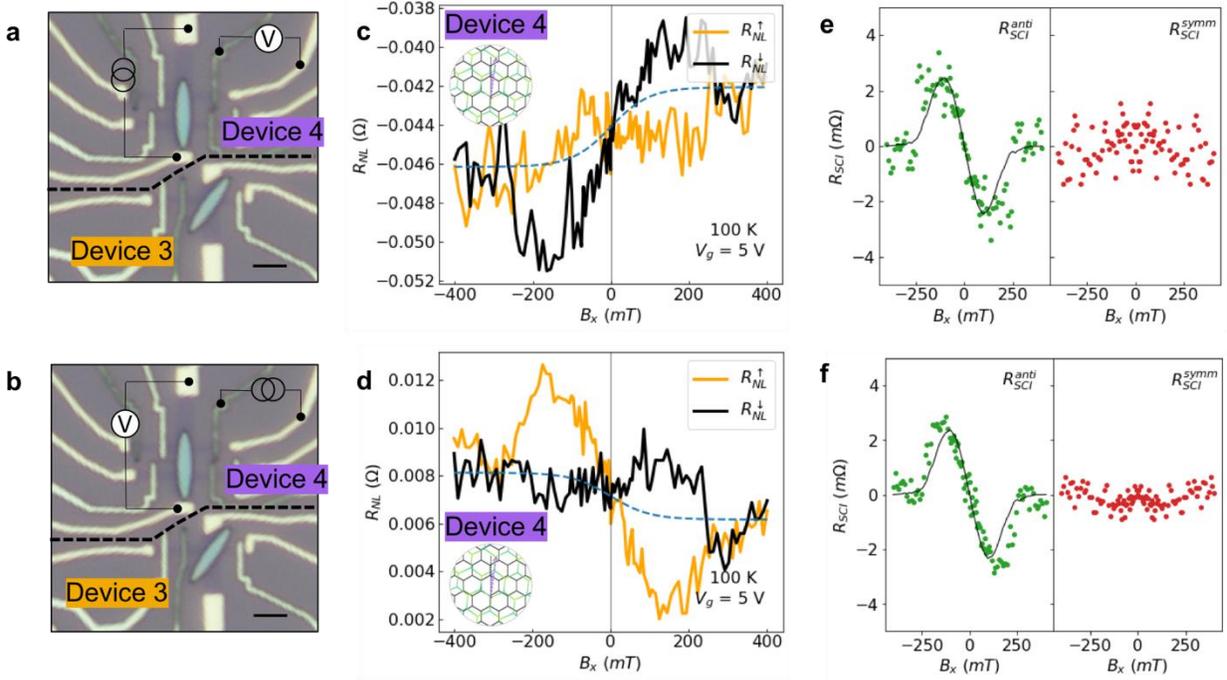

**Fig. S6 (a,b)** Measurement configuration of **(a)** charge-to-spin conversion and **(b)** spin-to-charge conversion for Device 4. **(c)** Charge-to-spin and **(d)** spin-to-charge non-local resistance as a function of $B_x$ with initial positive ($R_{NL}^{\uparrow}$, yellow line) and negative ($R_{NL}^{\downarrow}$, black line) magnetization direction of the FM injector, measured at 100 K and $V_g = 5$ V. The blue dashed lines show the S-shape background originating from the REE. **(e,f)** Net antisymmetric (green dots) and symmetric (red dots) Hanle precession signals extracted from the two curves in **(c)** and **(d)** by taking $R_{SCI} = (R_{NL}^{\uparrow} - R_{NL}^{\downarrow})/2$ and then extracting the antisymmetric and symmetric components, which arise from SHE and UREE, respectively. Unlike the results for Device 3 shown in Fig. 4d of the main text, no evidence of UREE signal is observed. The amplitude of the effect is the same for the REE, UREE and SHE when changing from spin-to-charge conversion to charge-to-spin conversion. However, the sign of REE changes, while the SHE keeps the same. These results fulfill the Onsager reciprocal relations[1].

**Note 7. Spin-charge interconversion of Device 3 and Device 4 as a function of $B_x$.**

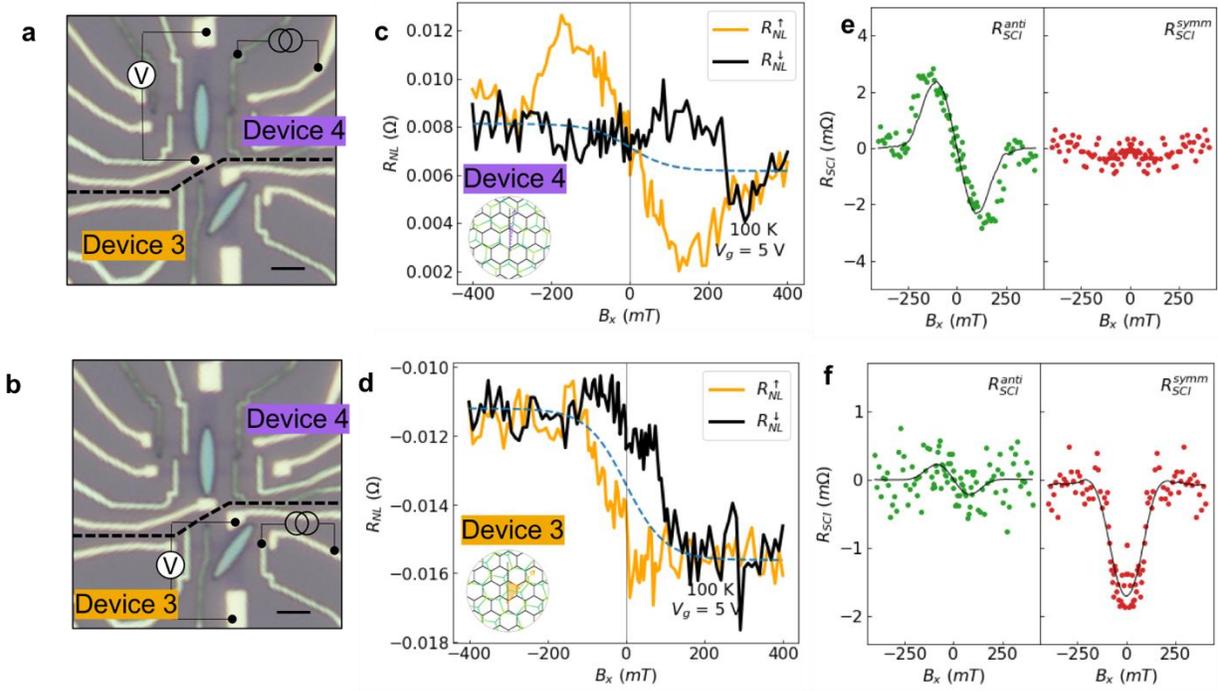

**Fig. S7 (a,b)** Measurement configuration of spin-to-charge conversion for **(a)** Device 4 and **(b)** Device 3. **(c, d)** Spin-to-charge non-local resistance as a function of $B_x$ with initial positive ($R_{NL}^{\uparrow}$, yellow line) and negative ($R_{NL}^{\downarrow}$, black line) magnetization direction of the FM injector, measured at 100 K and $V_g = 5$ V for **(c)** Device 4 and **(d)** Device 3. The blue dashed lines show the S-shape background originating from the REE. **(e,f)** Net antisymmetric (green dots) and symmetric (red dots) Hanle precession signals extracted from the two curves in **(c)** and **(d)** by taking $R_{SCI} = (R_{NL}^{\uparrow} - R_{NL}^{\downarrow})/2$ and then extracting the antisymmetric and symmetric components, which arise from SHE and UREE, respectively. Whereas a clear UREE signal is observed for Device 3 in **(f)**, no evidence of UREE signal is observed for Device 4 in **(e)**.

**Note 8. Spin-charge interconversion when changing the spin current direction.**

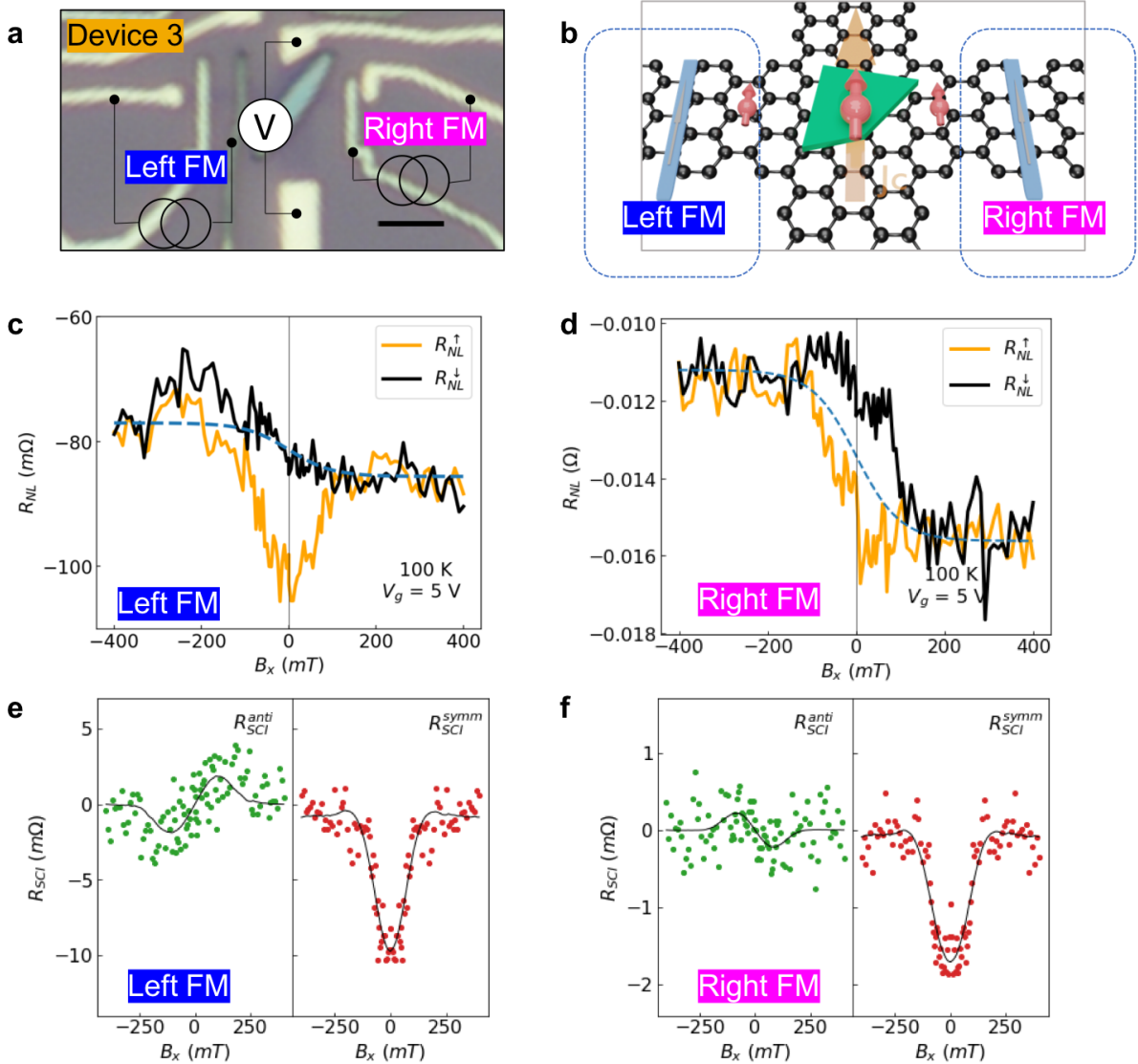

**Fig. S8** Spin-charge interconversion when changing the spin current direction by injecting the charge current from the other side of the Hall cross. The measurement schematic is shown in (**a**). Unlike SHE, the spin accumulation from UREE keeps the same for the left/right side in the transverse direction of the Hall cross, shown in (**b**). (**c,d**) Non-local resistance as a function of $B_x$ with initial positive ($R_{NL}^{\uparrow}$, yellow line) and negative ($R_{NL}^{\downarrow}$, black line) magnetization direction of (**c**) left FM and (**d**) right FM for Device 3. (**e,f**) Net antisymmetric (green dots) and symmetric (red dots) Hanle precession signals extracted from the two curves in (**c**) and (**d**) by taking $R_{SCI} = (R_{NL}^{\uparrow} - R_{NL}^{\downarrow})/2$ and then extracting the antisymmetric and symmetric components, which arise from SHE and UREE, respectively. Whereas the sign of SHE reverses when changing the spin current direction, it does not change for REE and UREE.

## Note 9. Spin-charge interconversion of Device 5.

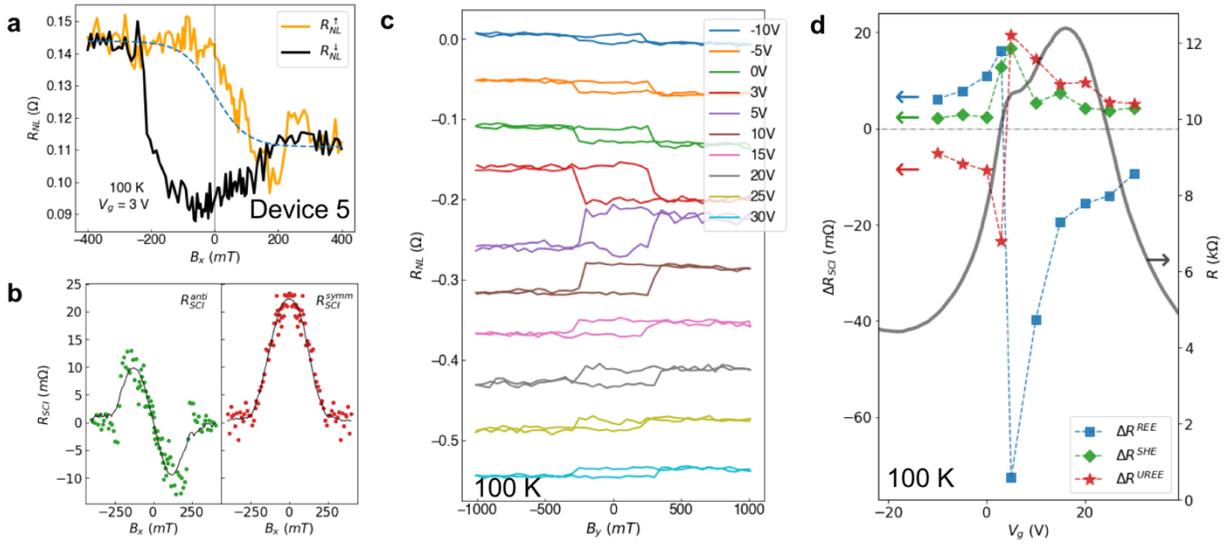

**Fig. S9.1 (a)** Non-local resistance as a function of $B_x$ with initial positive ($R_{NL}^{\uparrow}$, yellow line) and negative ($R_{NL}^{\downarrow}$, black line) magnetization direction of the FM injector, measured at 100 K and $V_g = 3$V for Device 5. The blue dashed line shows the S-shaped background originating from the REE. **(b)** Net antisymmetric (green dots) and symmetric (red dots) Hanle precession signals extracted from the two curves in **(a)** by taking $R_{SCI} = (R_{NL}^{\uparrow} - R_{NL}^{\downarrow})/2$ and then extracting the antisymmetric and symmetric components, which arise from the SHE and UREE, respectively. **(c)** Non-local resistance as a function of $B_y$ measured at 100 K and different $V_g$ for Device 5. **(d)** Gate-dependent SCI amplitude originating from REE (blue squares), SHE (green diamonds), and UREE (red stars) for Device 5. The grey solid line represents the two-point resistance of the Hall arm at 100 K.

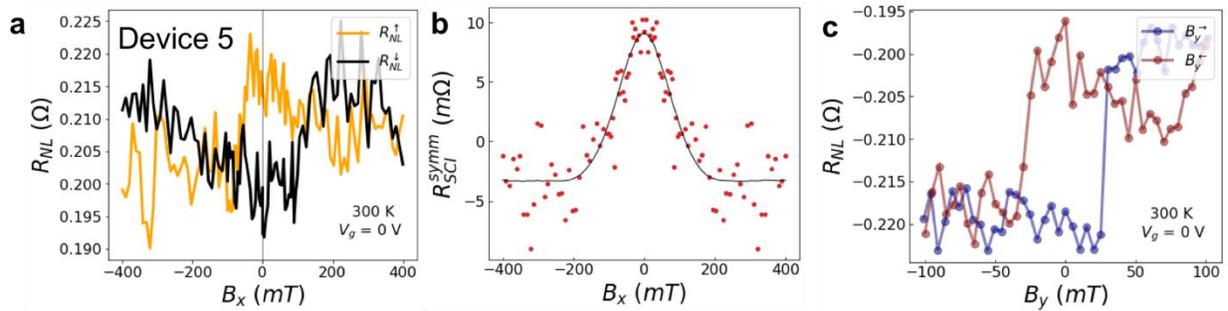

**Fig. S9.2 (a)** Non-local resistance as a function of $B_x$ with initial positive ($R_{NL}^{\uparrow}$, yellow line) and negative ($R_{NL}^{\downarrow}$, black line) magnetization direction of the FM injector for Device 5 measured at 300 K and $V_g = 0$ V. **(b)** Net symmetric Hanle precession signal extracted from the two curves in **(a)** by taking $R_{SCI} = (R_{NL}^{\uparrow} - R_{NL}^{\downarrow})/2$ and then extracting the symmetric component, which arises from the UREE. **(c)** Non-local resistance as a function of $B_y$ measured at 300 K and $V_g = 0$ V for Device 5.

**Note 10. Optical second Harmonic generation for Device 3 and Device 4.**

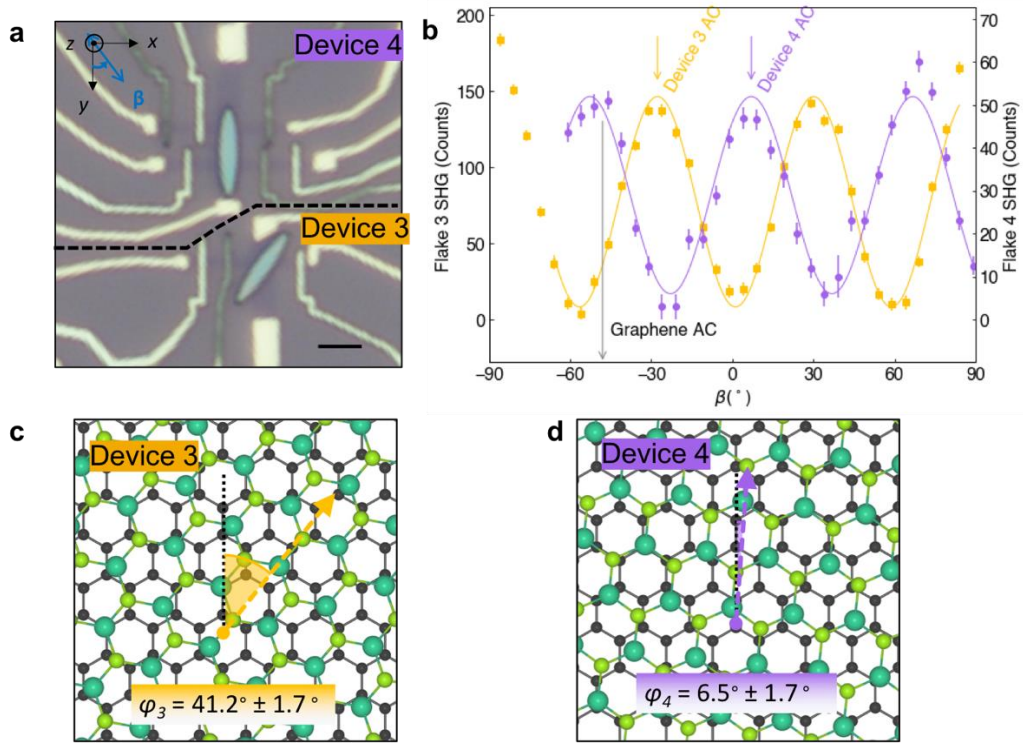

**Fig. S10 (a)** Optical image of Device 3 and Device 4. **(b)** SHG intensity as a function of polarization angle β for Flake 3 and Flake 4. **(c)** and **(d)** Moiré pattern of Device 3 and Device 4, showing the twist angle between graphene and WSe$_2$ of 41.2°±1.7 ° and 6.5 °±1.7 °, respectively.

**Note 11. Spin precession fitting process.**

**11.1 Symmetric Hanle fitting by considering low FM/graphene contact resistance.**

The contact resistance between FM and graphene in our devices is optimized between 1 kΩ and 3 kΩ for the low-noise non-local measurement and, therefore, the spin backflow effect at the interface should be taken into consideration. We model the spin diffusion as follows:

For the symmetric Hanle precession experiments obtained while having a parallel or antiparallel orientation of the FM magnetization, we model the spin diffusion in our devices using the Bloch equation:

$$D_s \nabla^2 \vec{\mu} - \frac{\vec{\mu}}{\tau_s} + \vec{\omega} \times \vec{\mu} = 0 \tag{1}$$

where $\vec{\mu}$ is the spin accumulation, $D_s$ the spin diffusion constant, and $\tau_s$ the spin lifetime. $\vec{\omega} = g\mu_B \vec{B}$ is the Larmor frequency, $g=2$ is the Landé factor, $\mu_B$ is the Bohr magneton, and $\vec{B}$ is the applied magnetic field.

The spin precession is induced in the $x - y$ plane when applying a magnetic field along $z$ direction. In this case, Eq. (1) turns into

$$D_s \frac{d^2}{dx^2} \begin{pmatrix} \mu_x \\ \mu_y \\ \mu_z \end{pmatrix} - \begin{pmatrix} \tau_{s\parallel}^{-1} \\ & \tau_{s\parallel}^{-1} \\ & & \tau_{s\perp}^{-1} \end{pmatrix} \begin{pmatrix} \mu_x \\ \mu_y \\ \mu_z \end{pmatrix} + \vec{\omega} \begin{pmatrix} \mu_x \\ -\mu_y \\ 0 \end{pmatrix} = 0 \tag{2}$$

The solution for $\mu_x$ and $\mu_y$ in Eq. (2) is

$$\mu_x = A e^{\frac{x}{\lambda_{s\parallel}}\sqrt{1+i\omega\tau_{s\parallel}}} + B e^{\frac{x}{\lambda_{s\parallel}}\sqrt{1-i\omega\tau_{s\parallel}}} + C e^{-\frac{x}{\lambda_{s\parallel}}\sqrt{1+i\omega\tau_{s\parallel}}} + D e^{-\frac{x}{\lambda_{s\parallel}}\sqrt{1-i\omega\tau_{s\parallel}}} \tag{3}$$

$$\mu_y = -iA e^{\frac{x}{\lambda_{s\parallel}}\sqrt{1+i\omega\tau_{s\parallel}}} + iB e^{\frac{x}{\lambda_{s\parallel}}\sqrt{1-i\omega\tau_{s\parallel}}} - iC e^{-\frac{x}{\lambda_{s\parallel}}\sqrt{1+i\omega\tau_{s\parallel}}} + iD e^{-\frac{x}{\lambda_{s\parallel}}\sqrt{1-i\omega\tau_{s\parallel}}} \tag{4}$$

where $\lambda_{s\parallel} = \sqrt{\tau_{s\parallel} D_s}$ is the spin diffusion length. $A$, $B$, $C$ and $D$ are the coefficients determined by the boundary conditions. The spin current is defined as:

$$I_{Sx(y)} = -\frac{W_{gr}}{eR_\square} \frac{d\mu_{x(y)}}{dx}, \tag{5}$$

where $W_{gr}$ is the width of the graphene channel and $R_\square$ is the square resistance of graphene. The spin accumulation at the detection position $x = x_{det}$ is then converted into a voltage with

$$V_{det}^S = P \frac{\mu_y(x_{det})}{e}, \tag{6}$$

where $P$ is the spin polarization of the Co detector. We consider that the Co injector has the same spin polarization. $V_{det}^s$ is usually normalized by the charge current $I_c$, giving the non-local resistance

$$R_{NL} = \frac{V_{det}^s}{I_c} = \pm R^s + R^{s0}. \tag{7}$$

$R^{s0}$ is a constant that corresponds to the spin signal at zero magnetic field. The net spin signal $\Delta R_{NL}$ is then expressed as:

$$\Delta R_{NL} = \frac{R_{NL}^P - R_{NL}^{AP}}{2} = R^s. \tag{8}$$

We thus determine the spin signal using the following boundary conditions:

1. Spin accumulation μ is continuous.

2. Spin current is continuous except:

    a. At F1 with spin injection by the charge current, by $\Delta I_s = I_c \cdot P/2$;

    b. At F1 and F2 due to the spin backflow effect because of the low contact resistance, by $\Delta I_s = -\mu_s/(2eR_c)$, where $R_c$ is the contact resistance.

3. Spin current is zero at the sample ends.

We write these boundary conditions in a matrix $X$ which fulfills: $MX = Y$, where $M$ contains the coefficients $M = A, B, C \ldots$ and $Y = (0, \ldots, \frac{I_c P}{2}, 0, \ldots)$ and use the Moore-Penrose inverse to invert $X^{-1}$ and obtain $M = YX^{-1}$. Using the numeric solution described above, we input the measured net spin signal $\Delta R_{NL}$, square resistance $R_\square$, channel size $W_{gr}$ determined by AFM, and the contact resistance $R_c$.

## 11.2 Spin precession fitting for SCI signals with out-of-plane magnetic field.

We determine the REE and UREE efficiency $\alpha_{REE(UREE)}$ and the in-plane spin lifetime $\tau_{s\parallel}^{WSe_2/gr}$ of the graphene in the heterostructure with the antisymmetric (symmetric) spin precession signal $\Delta R_{SCI}^{anti(symm)}$ originated from the REE (UREE) of proximitized graphene. The spin diffusion length corresponding to the in-plane spin lifetime is expressed as $\lambda_s^\parallel = \sqrt{\tau_{s\parallel}^{WSe_2/gr} D_s}$.

The non-local resistance $R_{SCI}^{anti}$ is given by

$$R_{SCI}^{anti} = \frac{\alpha_{REE} R_\square \overline{I_{Sx}}}{I_c} \tag{9}$$

where $\overline{I_{Sx}}$ is the average spin current in the cross-junction area of the Hall bar, which is defined as $\overline{I_{Sx}} = \frac{1}{W_{cr}} \int_L^{L+W_{cr}} I_{Sx}(x) dx$, where $L$ is the distance from F1 electrode to the edge of the Hall cross and $W_{cr}$ is the width of the cross. To determine $I_{Sx}(x)$, we use the equations and boundary conditions described in Section 11.1. We assume the spin diffusion coefficients for the TMDC-covered region and the pristine region to be equal for decreasing the number of variables.

Similarly, the non-local resistance $R_{SCI}^{symm}$ is given by

$$R_{SCI}^{symm} = \frac{\alpha_{UREE} R_\square \overline{I_{Sy}}}{I_c} \tag{10}$$

where $\overline{I_{Sy}}$ is the average spin current in the cross-junction area of the Hall bar. Here, we use the in-plane spin lifetime $\tau_{s\parallel}^{WSe_2/gr}$ result fitted from the REE curve, since the in-plane spin lifetime is considered to be equal for spins along $x$ and along $y$.

## 11.3 Spin precession fitting for spin and charge interconversion signals with in-plane magnetic field.

We determine the spin Hall angle $\theta_{SH}$ and the out-of-plane spin lifetime $\tau_{s\perp}^{WSe_2/gr}$ of the graphene in the heterostructure with the antisymmetric spin precession signal $R_{SCI}^{anti}$ originated from the (inverse) spin Hall effect of functionalized graphene. The spin diffusion length corresponding to the out-of-plane spin lifetime is expressed as $\lambda_s^\perp = \sqrt{\tau_{s\perp}^{WSe_2/gr} D_s}$. The non-local resistance $R_{SCI}^{anti}$ is given by

$$R_{SCI}^{anti} = \frac{\theta_{SH} R_\square \overline{I_{Sz}}}{I_c} \cdot \cos(\gamma) \tag{11}$$

where $\overline{I_{Sz}}$ is the average spin current in the cross-junction area of the Hall bar, which is defined as $\overline{I_{Sz}} = \frac{1}{W_{cr}} \int_L^{L+W_{cr}} I_{Sz}(x) dx$. In addition to the spin diffusion term, the $\cos(\gamma)$ indicates the non-local resistance affected by the contact pulling effect. Here, since the magnetic field is applied in-plane, the pulling effect for the FM here is relevant because of the small in-plane shape anisotropy of the FM electrodes. The magnetization is pulled by an angle $\gamma$ from their easy axis towards the field direction, injecting spins along $x$ and resulting in an additional term to the non-local resistance. Here $\gamma$ is extracted by measuring the symmetric Hanle precession with two FM with in-plane magnetic field, with details that can be found in our previous papers[2–4].

**Supplementary references**